\documentclass[useAMS,usenatbib,usegraphicx]{mn2e}
\usepackage{graphicx,times,epsfig,rotating}

\def\mnras{MNRAS}
\def\aj{AJ}
\def\aap{A\&A}
\def\apj{ApJ}
\def\apjl{ApJ}
\def\apjs{ApJS}
\def\araa{ARA\&A}
\def\pasp{PASP}
\def\nat{Nature}

\def\aaps{A\&AS}

\newcommand{\hpc}{\mbox{h$^{-1}$pc }}

\newcommand{\msun}{\mbox{M$_{\sun}$ }}
\newcommand{\msunend}{\mbox{M$_{\sun}$}}

\newcommand{\lco}{\mbox{L$_{\rm CO}$}}

\newcommand{\lmol}{\mbox{L$_{\rm mol}$}}

\newcommand{\cmthree}{\mbox{cm$^{-3}$}}

\newcommand{\msunyr}{\mbox{M$_{\sun}$yr$^{-1}$ }}

\newcommand{\htwo}{\mbox{H$_2$}}

\newcommand{\z}{\mbox{$z$}}

\newcommand{\zsim}{\mbox{$z\sim$ }}

\newcommand{\sunrise}{\mbox{\sc sunrise}}
\newcommand{\gadget}{\mbox{\sc gadget-3}}

\newcommand{\turtlebeach}{\mbox{\sc turtlebeach}}
\newcommand{\bzk}{\mbox{{\it BzK}}}

\newcommand{\sfrco}{\mbox{SFR $\propto {\rm CO}^{\alpha}$}}
\newcommand{\schmidt}{\mbox{SFR $\propto \rho^N$}}
\newcommand{\sigmacojone}{\mbox{$\Sigma_{\rm SFR}-\Sigma_{\rm CO J=1-0}^\alpha$}}
\newcommand{\sigmacojtwo}{\mbox{$\Sigma_{\rm SFR}-\Sigma_{\rm CO J=2-1}^\alpha$}}
\newcommand{\sigmacojthree}{\mbox{$\Sigma_{\rm SFR}-\Sigma_{\rm CO J=3-2}^\alpha$}}
\newcommand{\sigmaco}{\mbox{$\Sigma_{\rm SFR}-\Sigma_{\rm CO}^\alpha$}}
\newcommand{\sigmahtwo}{\mbox{$\Sigma_{\rm SFR}-\Sigma_{\rm H2}^\alpha$}}

\title[The KS Law at High Redshift]{The Kennicutt-Schmidt Star
  Formation Relation at \zsim 2}

\author[Narayanan, Cox, Hayward \& Hernquist]{Desika\, Narayanan$^{1}$\thanks{E-mail:
    dnarayanan@cfa.harvard.edu}\thanks{CfA
    Fellow}, Thomas
  J. Cox$^{2}$\thanks{Carnegie Fellow}, Christopher C. Hayward$^1$,
  Lars Hernquist$^1$\\$^{1}$Harvard-Smithsonian Center for
  Astrophysics, 60 Garden St., Cambridge, Ma 02138\\$^2$Observatories
of the Carnegie Institution of Washington, 813 Santa Barbara St.,
Pasadena, Ca, 91101}
\begin{document}

\date{MNRAS, 412, 287}

\pagerange{\pageref{firstpage}--\pageref{lastpage}} \pubyear{2010}

\maketitle

\label{firstpage}

\begin{abstract}
%Recent observations of excited CO emission lines from ``normal'' \zsim
%2 disc galaxies have shed light on the \sigmahtwo \ relation at
%high-\z \ via an observed \sigmacojthree \ relation.  Here, we
%describe a novel methodology for utilising data of this sort to
%understand the relationship between the observed CO emission lines in
%high-\z \ galaxies and the underlying SFR-$\rho^{N}$ Kennicutt-Schmidt
%relationship.  The main potential issue in interpreting the observed
%high-excitation

Recent observations of excited CO emission lines from \zsim 2 disc
galaxies have shed light on the \schmidt \ relation at high-\z \ via
observed \sigmacojtwo \ and \sigmacojthree \ relations.  Here, we
describe a novel methodology for utilising these observations of
high-excitation CO to derive the underlying Schmidt (\schmidt)
relationship.  To do this requires an understanding of the potential
effects of differential CO excitation with SFR.  If the most heavily
star-forming galaxies have a larger fraction of their gas in highly
excited CO states than the lower SFR galaxies, then the observed
molecular Kennicutt-Schmidt index, $\alpha$, will be less than the
underlying \schmidt \ index, $N$. Utilising a combination of SPH
models of galaxy evolution and molecular line radiative transfer, we
present the first calculations of CO excitation in \zsim 2 disc
galaxies with the aim of developing a mapping between various observed
\sigmaco \ relationships and the underlying \schmidt relation.  We
find that even in relatively luminous \zsim 2 discs, differential
excitation does indeed exist, resulting in $\alpha < N$ for highly
excited CO lines. This means that an observed (e.g.) \sigmacojthree
\ relation does not map linearly to a \sigmahtwo \ relation. We
utilise our model results to provide a mapping from $\alpha$ to $N$
for the range of Schmidt indices $N=1-2$.  By comparing to recent
observational surveys, we find that the observed \sigmacojtwo \ and
\sigmacojthree \ relations suggest that an underlying SFR $\propto
\rho^{1.5}$ relation describes \zsim 2 disc galaxies.

\end{abstract}

\section{Introduction}
\label{section:introduction}

Since the original works by \citet{sch59} and \citet{ken98b}
parameterising star formation rates (SFRs) in galaxies in terms of the
scaling relation:
\begin{equation}
\label{equation:schmidt}
SFR \propto \rho_{\rm gas}^{\rm N}
\end{equation}
there have been considerable efforts by both the Galactic and
extragalactic star formation communities to characterise the exponent
$N$ \citep[e.g. ][and references therein]{ken98a}. Constraining
the Schmidt SFR relation in galaxies is desireable, both for
understanding the physics of star formation on local scales, as well
as for giving simulators recipes for modeling processes below typical
numerical resolution scales.

Because the volume density is not observable, most observed forms of
Equation~\ref{equation:schmidt} have been in terms of the SFR and gas
surface densities\footnote{To remain consistent with typical
  literature nomenclature, we will refer to the volumetric form of
  Equation~\ref{equation:schmidt} as the Schmidt relation, and the
  surface density form as the Kennicutt-Schmidt (KS) relation.  We
  will reserve the index, $N$, for the volumetric exponent, and
  $\alpha$ for the surface density exponent.}.  As a result of recent
pioneering high resolution millimetre-wave surveys, two trends have become
apparent.

First, the SFR surface density in galaxies is well correlated with the
{\it molecular} gas surface density, though little relation exists
between the SFR and HI atomic gas \citep[e.g. ][]{big08,ler08,kru09}.
Second, the {\it global} integrated relationship between SFR and CO
molecular gas surface densities in observations of local galaxies
carries an exponent $N \sim 1.5$
\citep[e.g. ][]{san91,ken98b,gao04a,gao04b}.  A variety of theories
have been proposed to explain the $N \sim 1.5$ index inferred from
observations \citep[e.g. ][]{sil97,tan00,elm02a,kru05,kru09b}, most of
which rely on stars forming on a time scale proportional to the
dynamical time scale.

The recent advent of sensitive (sub)mm-wave interferometers has
allowed, for the first time, observational constraints on the \schmidt
\ relation in ``normal'' star-forming (e.g. not starbursting) galaxies
at high redshift via the detection of CO rotational emission lines
\citep[][]{bou07,ion09,dad10a,dad10b,bot09,gen10,tac10}.  However,
owing to fact that most mm-wave interferometers operate in the 1-3 mm
wavelength regime, most detections of CO emission lines at
high-redshift are of relatively excited lines (e.g. CO (J=3-2), as
opposed to the ground state transition which is typically used as a
tracer of \htwo \ molecular gas.

In principle, if the excitation of CO is relatively invariant in a
given sample of galaxies, one can simply make an assumption regarding
the CO (J=3-2) to (J=1-0) scaling ratio in high-\z \ galaxies, and
derive an \htwo \ gas mass (or surface density) with the inferred CO
(J=1-0) intensity.  However, if the excitation of CO in galaxies is
not constant with increasing SFR, then the exponent in the \sfrco
\ relation may not naturally translate to an exponent in the \schmidt
\ relation.  In other words, $\alpha$ does not always equal $N$ when
probing high excitation CO lines.

To see this, consider a sample of galaxies which are forming stars at a
rate according to SFR $\propto \rho^{1.5}$.  If the CO (J=1-0) line
faithfully traces the \htwo \ gas mass, then one could expect an
observed relationship $\Sigma_{\rm SFR} \propto \Sigma_{\rm CO
  J=1-0}^{1.5}$ \citep{kru07,nar08b}.  However, if the galaxies with
the highest SFR have a higher fraction of their CO gas excited into
the CO J=3 state than the lowest SFR galaxies, one will observe a
{\it flatter} relationship between SFR and L$_{\rm CO 3-2}$ than index
$\alpha=1.5$.  The \sigmacojthree \ relation will only map linearly to
the \sigmahtwo \ relation if the excitation of CO is invariant with
SFR.  In practice, this can only happen if the CO gas is thermalised
in the observed lines (if a given excited state is in LTE).

This effect of differential molecular excitation on observed molecular
SFR scaling relations has been observed in the local Universe.  The
SFR-CO (J=1-0) relationship has an index $\alpha \approx 1.5$, while
the SFR-CO (J=3-2) relationship has a flatter index $\alpha \approx
0.9$ \citep{san91,yao03,nar05,ion09,bay09}.  Similarly, while the
SFR-HCN (J=1-0) relationship is linear in local galaxies, the SFR-HCN
(J=3-2) index is decidedly sublinear \citep[with index $\alpha \sim
  0.7$; ][]{gao04a,gao04b,bus08,gra08,jun09}.  Observations of
individual star-forming clumps (which are massive enough to host
stellar clusters) have been inconclusive regarding whether these
global trends extend to smaller scales \citep[][]{wu05,wu10}.

In the absence of a more direct tracer of \htwo \ gas than
high-excitation CO, the potential effects of differential molecular
excitation with SFR need to be quantified in order to derive the
underlying SFR relation in high-redshift galaxies.  The few multi-line
constraints of excitation in high-\z \ galaxies hints that CO may be
subthermally excited even in the most luminous \zsim 2 systems
\citep{wei07,dan09,car10,har10}.  This indicates that applying a
uniform mapping from (e.g.) CO (3-2) to CO (J=1-0) line intensities
will indeed be problematic. In this arena, numerical models can offer
guidance.

Our aim in this paper is to calculate the mapping of observed \sigmaco
\ relations of excited lines (e.g. CO J=2-1 and CO J=3-2) to an
underlying \schmidt \ relation controlling the star formation.  In
\citet{nar09,nar10b,nar10a} and \citet{hay10}, we have developed a merger-driven
model for the formation of high-redshift ULIRGs which shows reasonable
correspondence with observed SEDs, CO emission properties and number
counts (C. Hayward et al. in prep.).  Here, we utilise the (idealised)
progenitor disc galaxies of these model mergers to represent the
star-forming discs at high-\z \ typically observed in CO emission line
surveys \citep[e.g. ][]{tac10,dad10b}.  We combine these hydrodynamic
simulations of disc galaxies with 3D non-LTE molecular line radiative
transfer calculations in order to calculate the full statistical
equilibrium excitation properties of the molecules.  These methods
allow us to determine the differential excitation of CO of \zsim 2
disc galaxies with respect to SFR, and derive a mapping of an observed
molecular Kennicutt-Schmidt law (\sigmaco) to a \schmidt
\ relationship.

\section{Numerical Methods}
\label{section:methods}

%\begin{figure}
%\hspace{-1cm}
%\includegraphics[angle=90,scale=0.4]{sfr_mstar.ps}
%\caption{SFR-M* relation for our model galaxies (blue diamonds),
%  plotted in relation to \zsim 2 observations by \citet[][ black plus
%    signs]{dad07}.  Our model galaxies lie toward the massive end of
%  SFR-M* relation, consistent with the typical galaxies studied for CO
%  emission at \zsim 2 \citep{dad10b,gen10}.\label{figure:sfr_mstar}}
%\end{figure}

Generally, our goal is to simulate galaxies hydrodynamically that
serve as reasonable representations of the star forming galaxies at
\zsim 2 residing on the ``main sequence'' of the SFR-M$^*$ relation
\cite[e.g. not starbursts; ][]{noe07b,noe07a}.  We first describe the
hydrodynamic methods employed, and then follow with our parameter
choices which ensure the physical properties of the model galaxies are
comparable to those observed.

We simulate the hydrodynamic evolution of the gas phase of our model
galaxies utilising the fully entropy-conserving $N$-body/SPH code
\gadget \ \citep{spr05b}.  The main components of the code relevant to
this study are the ISM and star formation prescriptions.

The ISM is modeled as multiphase in nature, with cold clouds embedded
in a hotter, pressure-confining ISM \citep[e.g. ][]{mck77,
  spr03a}. This ISM is pressurised against runaway fragmentation by
supernovae which is handled via an effective equation of state (EOS).
For details regarding this EOS, see Figure 4 of \citet{spr05a}.  Here,
we assume the stiffest equation of state \citep[$q_{\rm EOS} = 1$ in
][]{spr05a}.  Because the global excitation of CO is to some degree
dependent on the density structure of the ISM in the galaxy, we
explore relaxing this assumption with test cases of $q_{\rm EOS} =
0.75,0.25$.  We discuss the magnitude of uncertainty this parameter
causes in the next section.

Star formation is controlled by a volumetric Schmidt-relation such
that SFR $\propto \rho^{N}$.  In order to explore the effects of this
index on the observed SFR-\lmol \ relations, we have run simulations
varying $N$ between 1, 1.5 and 2.  We choose these values as
representative of the typical range of observed Kennicutt-Schmidt
indices at high-\z \ \citep[e.g. ][]{bou07,bot09,dad10b,gen10}. The
normalisation is anchored such that the a volumetric Schmidt relation
with index $N=1.5$ returns a surface-density Kennicutt-Schmidt
relation consistent with observations \citep{ken98b}.  The
normalisation of the $N=1$ and $2$ volumetric relations is forced to
match the $N=1$ relation at 20 times the density threshold for star
formation. This normalisation was chosen such that a disc modeled
after Milky Way parameters would have a global SFR of $\sim 2$ \msunyr
for all three cases. As we will discuss later, we explore the effects
of varying this normalisation by a factor of 2 in either direction.

It is worthwhile to consider that we force the stars to form according
to a volumetric Schmidt relation, though a surface density
Kennicutt-Schmidt relation is what is observable.  \citet{spr00} and
\citet{cox06a} have shown that a given Schmidt (volumetric) relation
and Kennicutt-Schmidt (surface density) relation have the same
exponent in disc simulations very similar to these, and we have
confirmed this equivalence with the simulations employed here.  This
similarity is not obvious.  \citet{sch08} have shown that a linear
mapping between Schmidt and Kennicutt-Schmidt SFR relations in
numerical simulations is dependent on the choice of equation of state.
The equation of state we employ (particularly the stiffest one,
$q_{\rm EOS} = 1$) is quite similar to  \citet{sch08}'s
``preferred'' EOS which reproduces the linear mapping between the
underlying Schmidt relation and observed Kennicutt-Schmidt relation.
We therefore refer to the volumetric and surface density relations
interchangeably\footnote{It is conceivable that in nature these
  relations are not in fact equivalent.  However, as this is presently
  unconstrainable, we are forced to operate under the assumption that
  the volumetric and surface-density exponents are indeed equivalent
  in real galaxies.}.

The simulations are not cosmological.  We simulate idealised discs in
order to maximize spatial resolution (here, the gravitational
softening length was 100 \hpc for baryons and 200 \hpc for dark
matter).  Because the simulations are not cosmological, we neglect
potential gas replenishment from the IGM.  However, we are less
concerned with the temporal evolution of the model galaxy, but rather
we aim to have the galaxy pass through phases during which it has
physical parameters comparable to those inferred at \zsim 2.  We thus
initialise our simulations with $f_g$=0.8, and allow the disc to
stabilise for some time.  We then consider the snapshots in our model
discs which have gas fractions between $f_g$=0.4-0.2 as motivated both
by measurements of \zsim 2 galaxies of this baryonic mass
\citep[c.f. \S~\ref{section:uncertanties};][]{erb06,dad10a,tac10}, as
well as the typical steady-state gas fraction of galaxies above
$M_{\rm bar} > 10^{11}$ in cosmological simulations
\citep[e.g. ][]{dav10}.  In a cases where the gas fraction remains
above $f_g > 0.2$, though the bulk of the gas is below the star
formation threshold, we arbitrarily cut out snapshots below SFR $< 5
$\msunyr to remain consistent with the SFRs of observed galaxies at
\zsim 2 \citep{dad10b}.

We model discs inside dark matter halos with \citet{her90} density
distributions of mass $\sim 3 \times 10^{12}$ \msun.  The haloes
are populated with discs who are constructed according to the
\citet{mo98} formalism.  These galaxies are bulgeless and have a total
baryonic mass of $\sim 2 \times 10^{11}$ \msunend, comparable to massive \bzk
\ galaxies at \zsim 2 \citep[e.g. ][]{dad07,tac10}. 

To summarise the galaxy evolution modeling, the galaxy snapshots
utilised in this study are ``selected'' according to particular
criteria in order to represent massive high-\z \ discs.  In
particular, the modeled galaxies have baryonic masses $M_{\rm bar}
\approx 2 \times 10^{11}$\msun and gas fractions $f_{\rm g}=0.2-0.4$.  When
analysing the synthetic SEDs from these galaxies, the modeled {\it
  BzK} colors\footnote{The synthetic colors are calculated with the
  dust radiative transfer code \sunrise \ \citep{jon06a,jon10a}.  We
  refrain from describing the parameters for the dust modeling as
  reporting the results from these calculations is not the primary
  objective of this paper. The \sunrise \ parameters utilised are
  identical to those in \citet{nar10a}, and we refer the interested
  reader to that paper for more details.  } are consistent with
selection as a star-forming {\it BzK} galaxy \citep[e.g. ][]{dad05}.
The result of this is a galaxy sample which lies on the high-mass end
of the \zsim 2 SFR-M* relation \citep{dad07}, consistent with the
typical galaxies observed for CO emission
\citep[e.g. ][]{dad10b,gen10}.  This selection returns roughly
$\sim$30-40 galaxies each for the $N = 1,1.5$ and $2$ model types.

The simulation snapshots are analysed in post-processing with
\turtlebeach \ in order to calculate their synthetic CO emission
properties.  \turtlebeach \ is a 3D non-LTE molecular line radiative
transfer code which considers both radiative and collisional
(de)excitation in calculating the excitation conditions
\citep{nar06b,nar08a_let} based on the \citet{ber79} method.

The process of calculating the excitation conditions in the molecular
gas is as follows.  First, the level populations (e.g. number of
molecules at a given excitation level) are guessed at. Based on this,
each molecular cloud in the galaxy emits CO line photons
isotropically, with directions determined via Monte Carlo draws.
These photons are absorbed by molecules in neighbouring clouds.  Once
the mean intensity, $J_\nu$, is known across the grid, the collisional
rates\footnote{The rates come from the {\it Leiden Atomic and
    Molecular Database} \citep{sch05}.} of CO with \htwo \ (based on
the gas densities) are calculated, and the level populations are
updated.  At this point, model photons are re-emitted based on the new
excitation conditions, and the whole process is repeated.  This
process is iterated upon until the level populations are converged.
%Formally, we consider convergence when the level populations in every
%cell are converged to 5\% (though note that typically the gas in the
%densest cells which contribute the bulk of the luminosity are
%converged to better than 1\%).

We assume that half of the cold gas in the \gadget \ simulations is in
molecular form \citep[as motivated by observations of local galaxies;
][]{ker03}.  By assuming that a constant fraction of the star-forming
gas in the \gadget \ simulations is molecular, we are effectively
assuming that the Schmidt law we impose in the SPH simulations is a
{\it molecular} Schmidt law, which we expect is reasonable.  Within
the \htwo \ gas, we assume the CO has a uniform Galactic abundance of
$1.5 \times 10^{-4} / \htwo $ \citep{lee96}.

The SPH simulations have resolution of $\sim$100 pc; as such, we do
not have information regarding the state of the \htwo \ gas below
these scales.  Thus, subgrid techniques are necessary to account for
the density distributions of molecular gas.  We assume the \htwo \ gas
in each cell is bound in giant molecular clouds (GMCs) with masses
randomly drawn from the Galactic mass spectrum \citep{bli07}, and
density distributions following power-law spheres. The clouds follow a
Galactic mass-radius relationship \citep[e.g. ][]{sol87,ros05,ros07}.
Together, these parameters set the density distribution of \htwo \ gas
in the model galaxies.  Because the excitation of molecules is
sensitive to the density distribution of gas in the galaxies, we
explore the effect of these parameters on our final results.
Observations of GMCs suggest a range of power-law indices, ranging
from $n=1$ to $2$ \citep{and96,ful92,wal90}.  Similarly, the index on
the GMC mass spectrum is thought to vary from $\gamma \approx -1.4$ to
$-2.8$ \citep[e.g. ][]{elm02b}.  Tests performed by \citet{nar08b}
have shown that within these ranges, the CO excitation in galaxies
similar to those presented here is not sensitive to choice of
mass-spectrum index or cloud power-law index.  Nominally, we employ
$n=1.5$ for the cloud power-law index and $\gamma = -1.8$.  A more
detailed description (including the underlying equations) regarding
our subgrid methods can be found in \citet{nar08a_let}.  Finally,
we have benchmarked our codes against literature standards
\citep{van02}, and published the results in \citet{nar06b}.

\section{Results and Application to Existing Observations}

\begin{figure}
\hspace{-1cm}
\includegraphics[angle=90,scale=0.4]{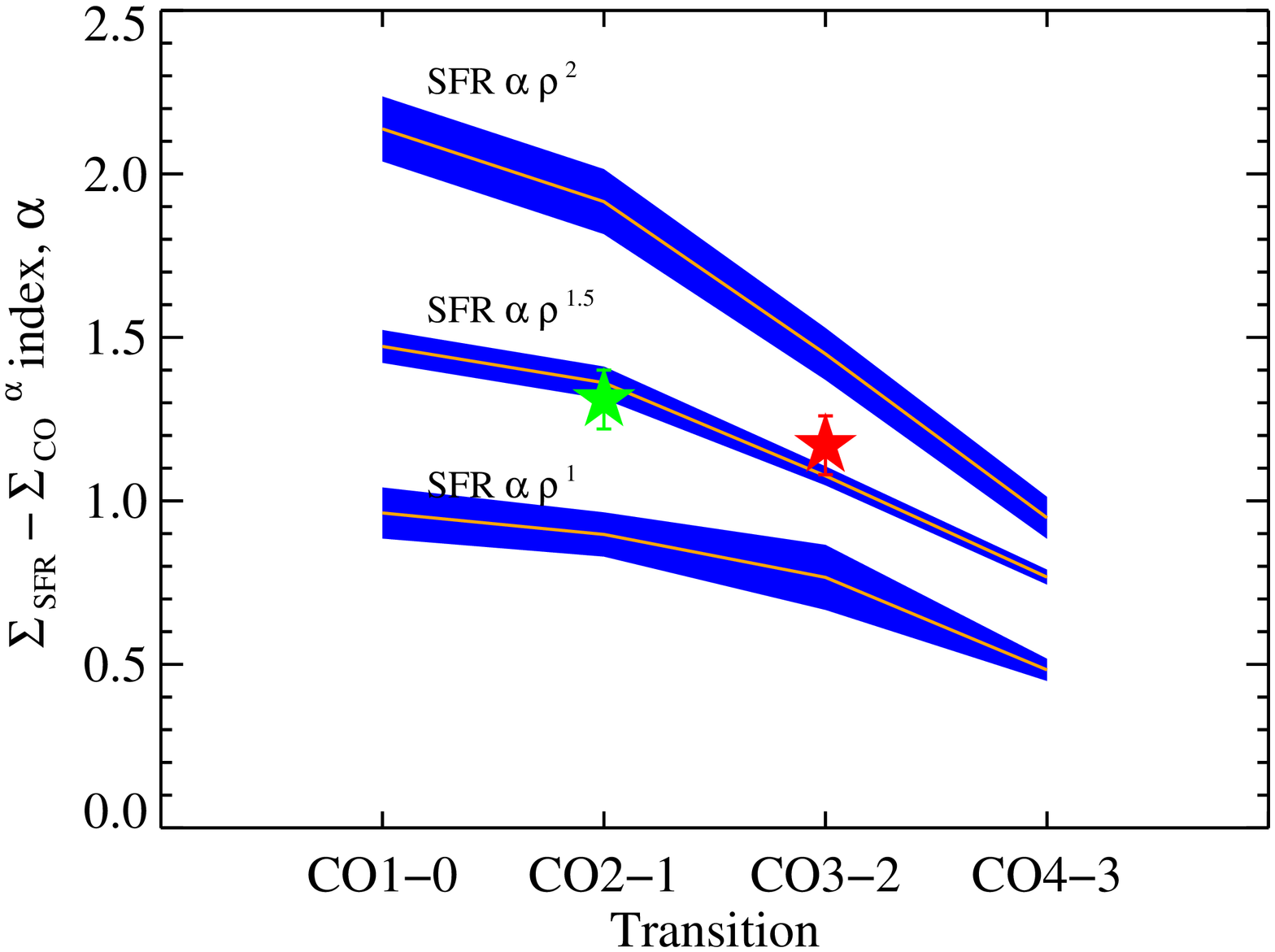}
\caption{Predicted \sigmaco \ index, $\alpha$, as a function of
  molecular transition for three different galaxy models.  The
  different galaxy models vary the relation that controls their SFR
  such that SFR $\propto \rho^N$ where $N=1,1.5$ and $2$. The blue
  shaded region denotes the uncertainties associated with limited
  sample sizes.  Generally, the \sigmacojone \ relation serves as a
  reasonable proxy for the underlying Schmidt SFR relation.  With
  higher-lying lines, differential excitation of the molecule with SFR
  becomes an important effect, and the \sigmaco \ relation is
  flattened from the underlying \schmidt \ relation.  For high
  critical density tracers, $\alpha \neq N$. The green star shows the
  recent CO (J=2-1) data from \citet{dad10b}, and the red star shows
  the CO (J=3-2) measurements of \zsim 2 galaxies by \citet{gen10}.
  This data suggests an underlying Schmidt index of
  $N=1.5$. \label{figure:sfr_lmol}}
\end{figure}

\subsection{Main Results}
\label{section:mainresults}
In Figure~\ref{figure:sfr_lmol}, we plot the model \sigmaco \ index,
$\alpha$, as a function of CO transition for three model disc
galaxies.  The disc galaxies form stars according to varying Schmidt
laws, SFR $\propto \rho^{N}$, where $N = 1, 1.5$ and $2$.  The top
curve denotes the $N=2$ case, middle $N=1.5$, and bottom $N=1$.  The
indices, $\alpha$, are derived by fitting log(SFR) versus log($I_{\rm
  CO}$) for a random draw of 20 galaxies (corresponding to typical
sample sizes in the current literature) from our parent samples of
$\sim 30-40$. We do this 1000 times for each model and plot the
standard deviation in the derived $\alpha$-indices as the blue shaded
region. We note that there is an uncertainty of $\sim 25\%$ in all
mean values shown in Figure~\ref{figure:sfr_lmol} based on varying
initial conditions.  To avoid cluttering the figure, this uncertainty
is not denoted in Figure~\ref{figure:sfr_lmol} itself, though will be
discussed later when we provide a quantitative mapping between the
Schmidt and Kennicutt-Schmidt indices.  The bulk of the arguments made
in this paper are summarised in this figure.

First, we see that CO (J=1-0) generally serves a good tracer of the
H$_{\rm 2}$ \ molecular gas, and that the \sigmacojone \ relationship
maps reasonably well from the underlying SFR-$\rho^N$ relation.  This
is because CO (J=1-0) has a relatively low critical density
\citep{eva99}, and most of the molecular gas emits the J=1-0 line.
While CO (J=1-0) is typically optically thick {\it within} GMCs, owing
to large velocity gradients it is optically thin on galaxy-wide
scales.  Thus so long as GMC properties do not vary strongly from
galaxy to galaxy \citep[as local measurements suggest;
][]{sol87,ros05,bli06,bli07}, and their mass and radius distributions
are relatively narrow, then an increase in \htwo \ surface density
will correspond to an increase in the number of GMCs, and a
commensurate increase in CO (J=1-0) \ emission.  For this reason, the
\sigmacojone \ relation serves as a reasonably good proxy for the
Schmidt relation.

The situation is different for higher excitation CO emission lines
(e.g. CO J=3-2).  Higher lying lines have relatively high critical
densities.  For example, the CO (J=3-2) line requires typical
densities of $n \ga 10^4 $\cmthree \ to excite the line.  Because of
this, these lines are typically subthermal even in the most luminous
high-redshift galaxies
\citep[e.g. ][]{and00,pap02,gre03,hai06,wei07,dan09,car10,har10}, and
consequently do not trace the bulk of the molecular gas. For
higher-lying CO lines, the emission line luminosity increases {\it
  superlinearly} with increasing mean gas density (owing to the
combined effects of increased gas mass as well as increased
excitation), and the observed SFR-\lco$^\alpha$ index, $\alpha$, will
be less than the underlying Schmidt index, $N$
\citep{kru07,nar08b,jun09}.

Another way of saying this is that there is a differential excitation
for galaxies with increasing SFR (or mean gas density, $<n>$).  The
most heavily star-forming (densest) galaxies will have a higher (e.g.)
CO J=3-2/CO J=1-0 ratio than the lowest SFR galaxies.  Then, because
the SFR-CO (J=1-0) relationship traces the underlying Schmidt index,
$N$, the observed (e.g.) SFR-CO (3-2)$^{\alpha}$ relation for higher
lying lines will necessarily have a {\it flatter} relation than the
underlying Schmidt relation.  This trend will become more pronounced
as one observes increasingly higher CO (or any molecular) transitions
with higher critical densities.  This is shown explicitly in
Figure~\ref{figure:sfr_lmol}, where we see the SFR-\lmol$^\alpha$
index decrease as a function of increasing CO transition (or critical
density) for all model galaxies.  Moreover, this was shown to be an
observed phenomena in the local Universe in recent surveys by
\citet{bus08} and \citet{jun09}.

\begin{figure}
\hspace{-1cm}
\includegraphics[angle=90,scale=0.4]{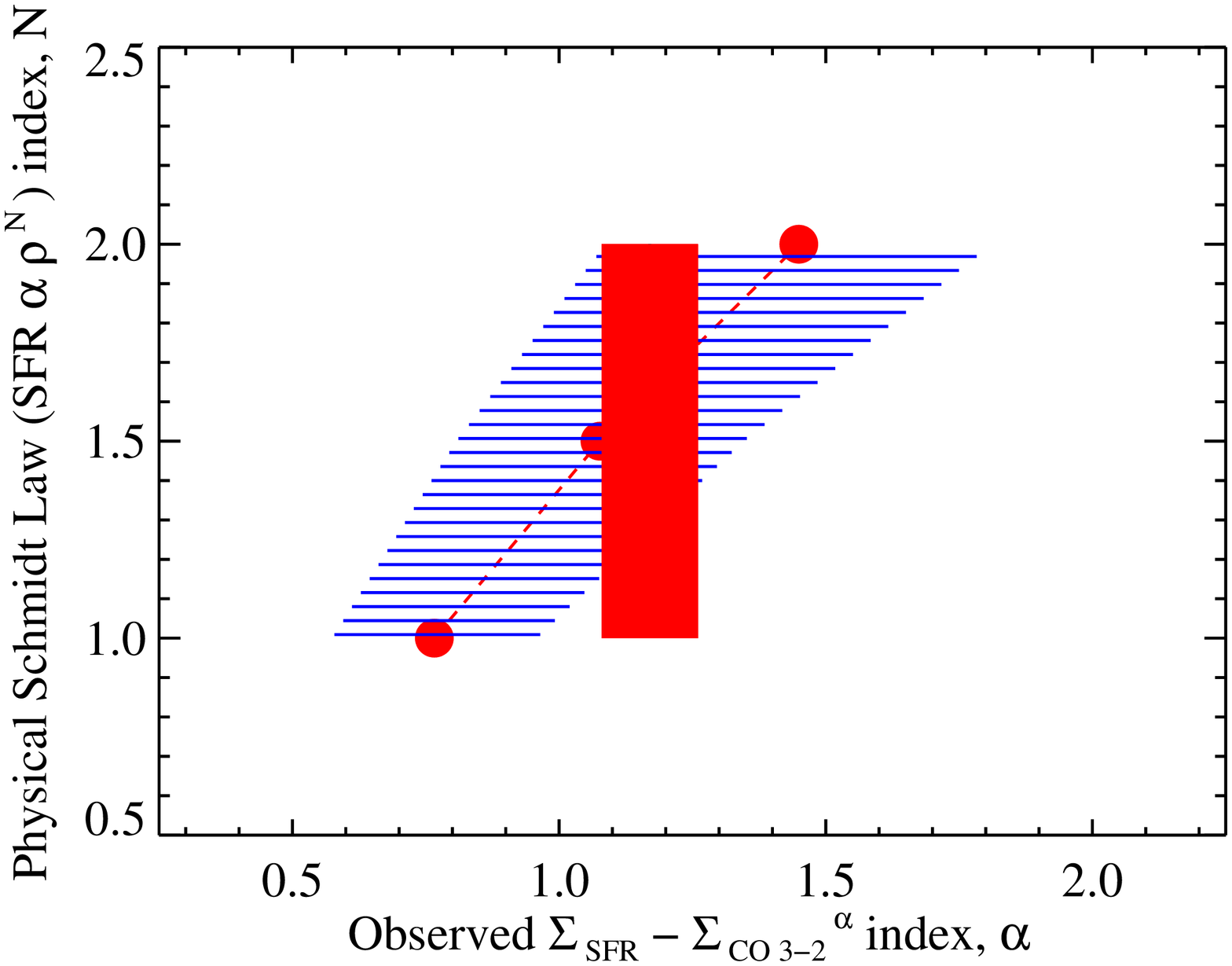}
\caption{The relationship between the \schmidt \ index, $N$, and the
  observed \sigmacojthree \ index, $\alpha$.  The red dashed line
  represents the mean result, and the blue shaded region denotes a
  25\% uncertainty (see text for details).  The red solid region
  represents the best fit to available CO (J=3-2) data from \zsim 2
  discs \citep{gen10}, and the thickness represents the associated
  uncertainty. The intersection of the observed \sigmacojthree
  \ index, $\alpha$ and the model results suggests that a Schmidt
  index of $N \approx 1.5$ may appropriately describe \zsim 2 disc
  galaxies.\label{figure:ks_molslope}}
\end{figure}

Figure~\ref{figure:sfr_lmol} therefore provides a mapping between the
\sigmaco \ relation for observed high-lying CO transitions and the
underlying Schmidt relation.  For the purposes of direct application
to current high-\z \ data, In Figure~\ref{figure:ks_molslope}, we turn
Figure~\ref{figure:sfr_lmol} into an actual mapping between the
\sigmaco \ index, $\alpha$, and the volumetric Schmidt index, $N$.  In
particular, we focus on the CO (J=3-2) transition as it is a
relatively commonly observed line at \zsim 2.  The solid line is the
mean from Figure~\ref{figure:sfr_lmol} (e.g. the mean after randomly
drawing 20 galaxies of our parent sample 1000 times).  The blue shaded
region denotes a 25\% range of uncertainty.  The uncertainty is
determined by characterising the dependence of $\alpha$ on the input
parameters. In particular, within the confines of our ``selection
criteria'', the input parameters which have the strongest effect on $\alpha$
are the equation of state and the Schmidt law normalisation (owing to
their changing the gas density distribution and the level of
thermalisation of the gas).  We find the maximum variance in $\alpha$
with these parameters is less than 25\%.

 We can utilise a combination of recent observations of high-\z
 \ star-forming systems and the information in
 Figures~\ref{figure:ks_molslope} and ~\ref{figure:sfr_lmol} to infer
 the Schmidt relation at high-\z.  Recent investigations by
 \citet{dad10b}, \citet{tac10} and \citet{gen10} have investigated the
 \sigmaco \ relation in ``normal'' star-forming systems at
 \z=1-2 which lie on the main sequence of the SFR-M$^*$ relation.
 \citet{dad10b} find a \sigmacojtwo \ index of $\alpha \approx 1.31$.
 Similarly, \citet{gen10} find a \sigmacojthree \ index for their data
 set of $\alpha \approx 1.17$.  Comparing this to
 Figure~\ref{figure:sfr_lmol}, we see that these values correspond to
 an underlying Schmidt index of $N=1.5$.

Figures~\ref{figure:sfr_lmol} and \ref{figure:ks_molslope} therefore
provide evidence that an Schmidt index $N \approx 1.5$ holds for
high-redshift galaxies.  It is important to note, however, that there
is an uncertainty of $\sim 25\%$ in these models.  Observations of
numerous CO transitions will help to narrow the uncertainty in the
derived \schmidt \ relation by providing additional constraints in the
$N-\alpha$ space probed in Figure~\ref{figure:ks_molslope}.  In
Table~\ref{table:indices}, we provide the mapping between Schmidt
indices, $N$ and molecular Kennicutt-Schmidt indices, $\alpha$ for
four CO transitions.  These numbers provide enough information to
create $N-\alpha$ plots similar to Figure~\ref{figure:ks_molslope} for
transitions other than CO (J=3-2), and test the concept that a Schmidt
index of $N \approx 1.5$ describes star formation in high-\z
\ galaxies.

\subsection{Uncertainties and Dependence on Model Physical Parameters}
\label{section:uncertanties}
\begin{figure}
\hspace{-1cm}
\includegraphics[angle=90,scale=0.4]{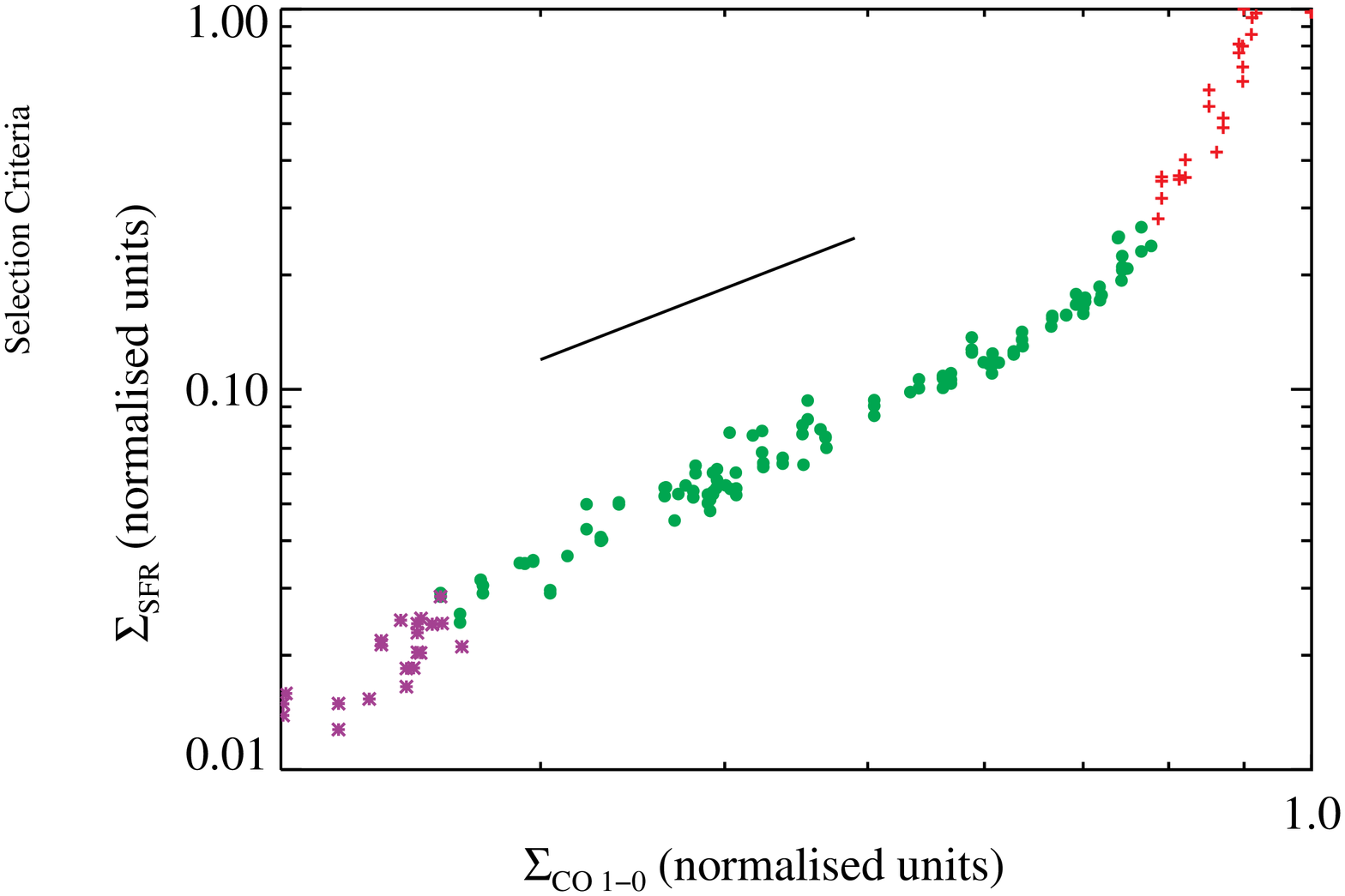}
\caption{The model \sigmacojone \ relation for our model with
  underlying Schmidt index $N=1.5$.  The solid line shows a slope of
  $N=1.5$. The ordinate and abscissa are in normalised units. The
  green circles show the galaxies that satisfy our ``selection''
  criteria culls.  Including extremely gas-rich (red plus signs) or
  gas-poor systems (where most of the gas is below the star formation
  threshold; purple crosses) may cause the observed KS indices to
  deviate from those in
  Figure~\ref{figure:sfr_lmol}\label{figure:ksplot}}
\end{figure}

 In Figure~\ref{figure:ksplot}, we show the simulated \sigmacojone
 \ plot for our model with a Schmidt index of $N=1.5$, and denote the
 galaxies which fall within our selection criteria.  The trends in
 Figure~\ref{figure:sfr_lmol} can depend on how large of a dynamic
 range the mock observations span.

The trends in Figure~\ref{figure:sfr_lmol} depend on the degree of
differential excitation, which depends on the dynamic range of SFR
surface densities (or gas surface densities) observed. For example,
consider the case where observations only probe a limited range of
extremely high SFR galaxies (here, we take this to mean galaxies with
gas fraction $0.4 < f_g < 0.8$).  In the high gas fraction/high SFR
regime, the excitation conditions vary extremely rapidly.  The most
gas-rich, densest systems have much of their gas thermalised, whereas
lower gas fraction galaxies in this range ($f_g \approx 0.4-0.6$)
begin to contain subthermal gas.  While it may seem counterintuitive
that some high gas-fraction galaxies may not be fully thermalised in
higher CO lines, it is important to remember that the simulations (as
do observations) consider {\it global} emission, from both low and
high-density regimes.  Subthermal CO level populations have been noted
both in observations of local galaxies \citep{nar05,nar08d,ion09}, as
well as in observations of even the densest, most heavily star-forming
submillimetre-galaxies at high-redshift \citep{car10,  har10}.  The rapidly
dropping CO excitation (from the higher J states to the ground states)
at the highest SFRs causes the observed molecular KS relation to
steepen from the underlying Schmidt relation.  In the fiducial case of
the $N=1.5$ galaxy, the observed \sigmacojone \ index can range from
$\sim$1.5 to $\sim$2.7 as we consider increasing numbers of galaxies
with gas fractions beyond $f_g > 0.4$ in the fits.

The molecular KS relation at the other end of the spectrum, in
gas-poor galaxies, is less clear.  On one hand, we can expect that it
would steepen owing to the bulk of the gas in the galaxy being below
the SF threshold \citep[see, e.g. ][]{big08}.  However, our models
don't consider the possible destruction of molecular gas in low
density environments \citep[e.g. ][]{kru09a}.  At low densities, the
neutral ISM is a mixture of HI and \htwo.  This is not captured by our
models as we are forced by lack of resolution to consider the \htwo
\ as a fixed fraction of the neutral ISM mass.  In reality, if the
\htwo \ and/or CO mass also drops at low densities, the observed KS
relation may not become as steep as Figure~\ref{figure:ksplot} would
suggest.

The trends predicted in Figure~\ref{figure:sfr_lmol} are relatively
robust within the physical parameter range chosen for the bulk of this
paper ($0.2 < f_g < 0.4$, with a typical dynamic range in SFR of order
$\sim$10).  As was just shown, large deviations from galaxies of this
sort may change the observed mapping from a Schmidt relation to a KS
relation.  That said, our assumed range of gas fractions may
accurately represent real \zsim 2 galaxies.  For example, cosmological
hydrodynamical simulations suggest that most galaxies with baryonic
masses $>$ a few $\times 10^{11}\msun$ (comparable both to those modeled, as
well as those observed) have steady-state gas fractions of order
$\sim$0.2-0.4 \citep[e.g. ][]{dav10}. Recent observations appear to
come to a similar conclusion.  For example, CO measurements of \bzk
\ galaxies by \citet{dad10a} suggest that the expected gas fraction
for galaxies of the baryonic mass modeled here are $f_g \sim 0.4$.
Similarly, by inversion of the Kennicutt-Schmidt relation,
\citet{erb06} suggest gas fractions of massive star-forming galaxies
of order $f_g \approx 0.2-0.4$.  Finally, a large CO survey by
\citet{tac10} suggest that the distribution of gas fractions of
normal, star-forming galaxies in this mass range between \z=1-3 is
relatively sharply peaked at $f_g = 0.3-0.4$ with only rarer
excursions outside of this range.

It is additionally worth discussing the potential effects of our ISM
assumptions on our modeled results.  As was discussed in
\S~\ref{section:methods}, the hydrodynamic galaxy evolution
calculations cannot resolve giant molecular clouds, the primary origin
site of CO emission lines.  Because of this, we are forced to utilise
a subgrid prescription for including GMCs in the radiative transfer
calculations.  In the absence of any observational constraints of the
nature of the structure of the molecular ISM at high-\z, we assume
that clouds exist as bound spheres, following a Galactic mass-spectrum
and Galactic mass-radius relation with a power-law density gradient
\citep[see ][for the actual underlying algorithms]{nar08a}.

Our model results are dependent on this assumption.  As was shown in
\citet{kru07} and \citet{nar08b}, the degree of differential
excitation in galaxies comes from the shape of the high-density tail
of the molecular gas density distribution.  While determining the
exact dependence of the molecular excitation on the density
distribution of molecular gas is worthy of an independent study,
previous model results suggest that a range of reasonable assumptions
result in similar differential excitation patterns.  For example,
\citet{kru07} assumed a log-normal density distribution in their model
galaxies, whereas \citet{nar08b} utilised a numerical model derived
from the aforementioned cloud mass spectrum, mass-radius relation, and
power-law density distribution within clouds.  Both studies found a
similar degree of differential excitation in local galaxies
\citep[see, e.g., the similar predictions for the SFR-HCN 3-2 relation
  in local galaxies between the two models in Figure 2 of
][]{bus08}. In this sense, our model assumptions for the structure of
the molecular ISM appears reasonable, at least in the context of local
galaxies\footnote{It is additionally worth noting that \citet{nar08b}
  varied the range of indices of the Galactic mass-spectrum and
  power-law density gradients within the range of observational
  constraints, and found little difference in the final differential
  excitation in their modeled local galaxies.}.  In a similar vein, we
note that the assumed equation of state in our models can affect the
density distribution via pressurisation of the ISM.  As noted in
\S~\ref{section:mainresults}, within the range of our modeled EOSs
\citep[$q_{\rm EOS}=0.25-1$, see ][and \S~\ref{section:methods} for
  more details]{spr05b}, we find $<25 \%$ difference in the modeled
molecular KS indices.

It is conceivable, however, that the molecular gas in high-redshift
galaxies is different in its structure than clouds in the Galaxy.  For
example, observations of local ULIRGs show that in high gravitational
potentials, tides may cause the molecular ISM to exist in a smooth
structure with a large volume-filling factor, rather than in bound
clouds \citep{dow98}.  It is not entirely clear, in these starburst
galaxies, whether the density distribution is similar in shape to
those in normal discs (e.g. with a high-density tail), or whether they
have a substantially different gas density distribution.  Resolved
maps of GMCs in local ULIRGs with ALMA will help to answer this
question.  

Finally, we note that galaxies at high-\z \ may have a larger fraction
of their gas in a dense phase than local galaxies.  This is explicitly
accounted for in our modeling.  As described in \citet{nar08a}, clouds
are randomly drawn from the mass spectrum to fill cells of a given
mass until the mass-budget is used.  When the density in a particular
region is high (as informed by the hydrodynamic models), more
high-mass clouds will statistically be drawn, thus increasing the
dense gas fraction.  

It is important to note, however, that the effects of radiative
transfer on scales below our cell-sizes are only marginally accounted
for in these models.  When a photon is emitted, it sees a column
density drawn randomly from the distribution of columns seen in the
cell \citep{nar08a}, and deposits some intensity.  However, because
the velocity distribution is not modeled on sub-grid scales, this is a
limiting assumption.  Some of this radiation may actually emerge from
the cell owing to large velocity gradients within a cell which may
keep photons from being trapped.  Efforts are underway to explicitly
account for this effect, though it is a task well outside the scope of
this study.

\section{Discussion}

\begin{table*}
\centering
%\begin{minipage}{100mm}
\caption{This Table provides the mapping between observed
  Kennicutt-Schmidt molecular surface-density indices, $\alpha$, and
  underlying volumetric Schmidt indices, $N$.  We provide the mapping
  for CO transitions J=1-0 through J=4-3.  The numbers contained
  here constitute the information necessary to re-create a plot like
  Figure~\ref{figure:ks_molslope} for any of the four modeled CO
  transitions and will aid interpretation of future observations of
  varying molecular transitions.  The ``errors'' denote a 25\%
  uncertainty level which encompasses variations in the final solution
  upon changing initial conditions in our simulation (primarily the
  equation of state and normalisation of the SFR relation).}
\begin{tabular}{@{}c||cccc@{}}
\hline Schmidt Index $N$ & SFR-CO (J=1-0)$^\alpha$ & SFR-CO
(J=2-1)$^\alpha$ & SFR-CO (J=3-2)$^\alpha$ & SFR-CO (J=4-3)$^\alpha$
\\ \hline

1  & 0.96$\pm 0.24$   &  0.89$\pm0.22$   &0.76$\pm 0.19$   &0.49$\pm 0.12$\\
1.5& 1.47 $\pm 0.37$  &  1.36$\pm0.34$   &1.08$\pm 0.27$   &0.77$\pm 0.19$\\
2&   2.13 $\pm 0.53$   &  1.95$\pm 0.48$  &1.45$\pm 0.36$   &0.95$\pm 0.24$\\

\hline
\end{tabular}
%\end{minipage}
\label{table:indices}
\end{table*}

The principle result of this study is that, due to subthermal
excitation in high-lying CO lines in high-\z \ galaxies, observed
\sigmaco \ relations do not necessarily map linearly to \sigmahtwo
\ relations.  This is part of a broader theoretical framework first
developed by \citet{kru07} and \citet{nar08b} which posits that the
underlying density distribution of galaxies is crucial in mapping the
underlying Schmidt relation to observed molecular Kennicutt-Schmidt
relations.  In this paper, we have expanded upon these studies by
exploring the relationship between Schmidt and Kennicutt-Schmidt
indices in models which aim to serve as reasonable representations of
the \zsim 2 star-forming disc galaxies being uncovered in sensitive
optical/NIR observations \citep[e.g. ][]{dad05,for09}.  Indeed, the
models studied here have been shown in previous publications to
satisfy the star-forming {\it BzK} colour-selection criteria
\citep{nar10b}, as well as serve as reasonable progenitors for
luminous, merger-driven \zsim 2 submillimetre and 24 \micron-selected
galaxies \citep{hay10,nar10a}.  We have additionally explored the
dependence of the observed molecular Kennicutt-Schmidt index on
varying underlying Schmidt relations. Our models, combined with the
general theoretical framework of \citet{kru07} and \citet{nar08b}
suggest that observed \zsim 2 discs are subject to differential CO
excitation with respect to SFR, and that the observed \sigmacojtwo and
\sigmacojthree \ relations may map to an underlying Schmidt index of
$N=1.5$ controlling the SFR.

An important verifying aspect to these models is that they are able to
explain the multitude of observed SFR-\lmol \ relations in the local
Universe.  For example, a linear relationship has been observed
between SFR and HCN (J=1-0) in local galaxies, a trend which has often
been interpreted as evidence that a linear {\it dense gas} Schmidt
relation controlled the SFR.  Our models suggest the linear SFR-HCN
(J=1-0) relation is in reality a combined effect of an underlying
\schmidt \ index of $N=1.5$ and differential excitation in HCN
\citep{kru07,nar08b}.  This view has been observationally confirmed by
\citet{bus08}, who showed that local galaxies exhibit an {\it
  sublinear} SFR-HCN (J=3-2) relation, and thus follow the trend of
decreasing SFR-HCN$^\alpha$ index with increasing transition number
characteristic of these models
\citep[e.g. Figure~\ref{figure:sfr_lmol}; see also Figure 7 of
][]{nar08b}.  Similarly, this model satisfies the multi-line
constraints of local CO observations.  The SFR-CO (J=1-0) relation in
local galaxies appears to have an index ranging from $\sim 1.3-1.5$.
At higher-lying transitions (e.g. CO J=3-2), the index drops to $\sim
0.9$, in accordance with theoretical predictions \citep{ion09}.  We
note that at lower bolometric luminosities, even the SFR-CO (J=1-0)
relation may be subject to differential excitation and serve as a
relatively poor tracer of the underlying Schmidt relation.

Finally, with an eye toward ALMA, we comment on the role of spatial
resolution in observational determinations of the KS relation at
high-\z.  The exact mapping between the observed \sigmaco \ index,
$\alpha$, and the \schmidt \ index, $N$, depends on the level of
thermalisation of the gas within the beam.  It is dependent (to first
order) on the mean gas density.  Higher resolution observations which
probe just the nucleus of the galaxy will probe higher mean densities,
and allow even higher-lying CO emission lines to directly trace the
underlying Schmidt SFR relation.  Indeed, some very tentative
observational evidence for this trend in local galaxies has been shown
by \citet{nar08d}. The trends shown in Figures~\ref{figure:sfr_lmol}
and \ref{figure:ks_molslope} are for the central 6 kpc of the galaxy,
comparable to the typical resolution of current interferometric
observations of \zsim 2 galaxies \citep[e.g. ][]{tac10}. Our models
suggest that observations of the central $\sim$2 kpc will probe
sufficiently dense gas that the observed \sigmacojthree \ index,
$\alpha$, will trace the underlying \schmidt \ index, $N$.

\section{Summary}

Current facilities demand that observations of the molecular
Kennicutt-Schmidt relation at \zsim 2 probe highly excited CO lines
(e.g. CO J=3-2).  However it is not clear how exactly to map observed
\sigmacojthree \ relations to underlying \schmidt \ relations:
differential excitation of CO with SFR may make interpretation
difficult.

In order to aid in the interpretation of observed molecular
Kennicutt-Schmidt relations, we have calculated the first models of CO
excitation for star-forming disc galaxies at \zsim 2.  Our main results are:

\begin{enumerate}

\item Due to differential excitation of CO with SFR in \zsim 2 disc
  galaxies, global observations of a (e.g.) \sigmacojthree \ relation
  will result in a {\it flatter} index, $\alpha$, than the underlying
  Schmidt (\schmidt) index, $N$. This trend
  exists for all lines above CO (J=1-0), though the disparity between
  $\alpha$ and $N$ grows with increasingly high CO transitions owing
  to increasing critical densities in the line.

\item We present a mapping from observed \sigmaco \ indices,
  $\alpha$, to underlying Schmidt indices, $N$ for global observations
  of \zsim 2 discs. Combining these model results with the observed (nearly)
  linear relationship between $\Sigma_{\rm SFR}$ and $\Sigma_{\rm CO
    J=3-2}$ suggests that a relation SFR $\propto \rho ^{1.5}$
 may  control the SFR in \zsim 2 galaxies.

\end{enumerate}

\section*{Acknowledgements}
 We thank Shane Bussmann, Neal Evans, Reinhard Genzel, Patrik Jonsson,
 Dusan Keres, Mark Krumholz, Charlie Lada, Kai Noeske, Alice Shapley,
 Amiel Sternberg and Linda Tacconi for enjoyable conversations as the
 ideas for this study were developed.  The simulations in this paper
 were run on the Odyssey cluster, supported by the Harvard FAS
 Research Computing Group.

\end{document}